\documentclass[a4paper,english,12pt]{article}

\usepackage{fontenc}
\usepackage[latin1]{inputenc}
\usepackage{float}
\usepackage{color}
\usepackage{graphicx}
\usepackage{amssymb}
\usepackage{babel}
\usepackage{appendix}
\usepackage{amsmath}
\usepackage{fancyhdr}
\usepackage[affil-it]{authblk}
\usepackage{geometry}
\title{Multi-faceted plasmonic nanocavities}
\author[1]{Kalun Bedingfield}
\author[1,2]{Eoin Elliott}
\author[1]{Arsenios Gisdakis}
\author[3,4]{Nuttawut Kongsuwan}
\author[2]{Jeremy J Baumberg\thanks{jjb12@cam.ac.uk}}
\author[1]{Angela Demetriadou\thanks{a.demetriadou@bham.ac.uk}}

\affil[1]{School of Physics and Astronomy, University of Birmingham, Edgbaston, Birmingham, B15 2TT, UK}
\affil[2]{NanoPhotonics Centre, Cavendish Laboratory, University of Cambridge, Cambridge CB3 0HE, UK}
\affil[3]{Thailand Center of Excellence in Physics, Ministry of Higher Education, Science, Research and Innovation, Bangkok 10400, Thailand}
\affil[4]{Quantum Technology Foundation (Thailand), Bangkok 10110, Thailand}

\date{Dated: \today}

\begin{document}

\maketitle

\begin{abstract}
Plasmonic nanocavities form very robust sub-nanometer gaps between nanometallic structures and confine light in deep subwavelength volumes to enable unprecedented control on light-matter interactions.
However, spherical nanoparticles acquire various polyhedral shapes during their synthesis, which has defining impact on controlling many light-matter interactions, such as photocatalytic reactions.  
Here, we focus on nanocavities made of three polyhedral nanoparticles (cuboctahedron, rhombicuboctahedron, decahedron) that commonly occur during the synthesis of spherical nanoparticles. Their photonic modes have a very intricate and rich optical behaviour, both in the near- and far-field.
Through a recombination technique, we obtain the total far-field produced by a molecule placed within these nanocavities, to reveal how energy couples in and out of the system. 
This work paves the way towards understanding and controlling light-matter interactions, such as photocatalytic reactions and non-linear vibrational pumping, in such extreme environments. 
\end{abstract}

\begin{figure}[b!]
\centering
\includegraphics[width=\linewidth]{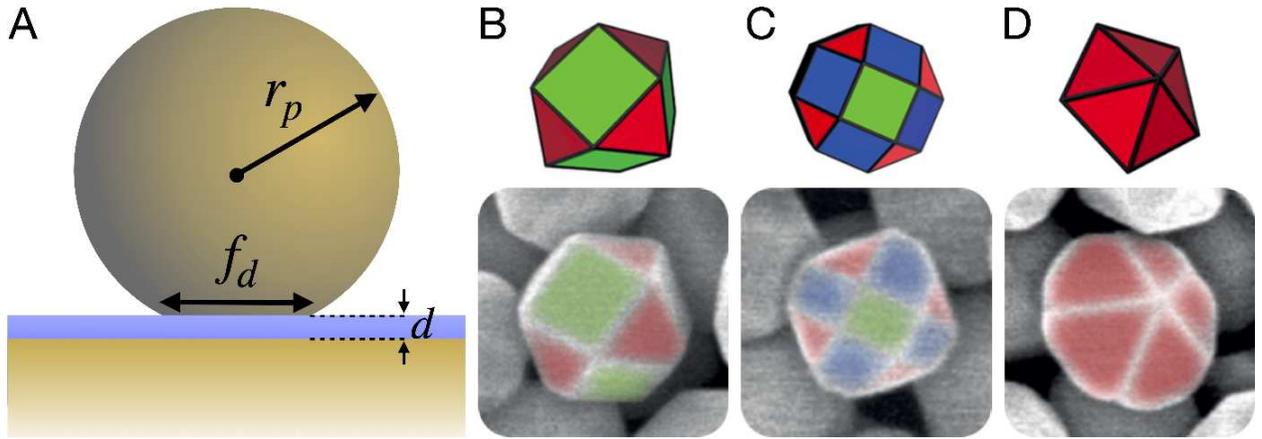}
\caption{\label{fig:polygons}A. TSoM schematic for a gold NP of radius $r_p$ with a circular facet of diameter $f_d$ assembled a distance $d$ above a flat gold substrate, separated by a spacer of refractive index $n$. Polyhedral NP structures above their corresponding scanning electron microscopy images: (B) cuboctahedron, (C) rhombicuboctahedron and (D) decahedron~\cite{Benz2016}. The green, red and blue facets refer to planes of gold atom crystallization $\{100\}$, $\{110\}$ and $\{111\}$, respectively.}
\end{figure}

Plasmonic nanocavities confine electromagnetic fields to extremely sub-wavelength volumes and massively enhance light-matter interactions. They are typically formed by bringing close together two or more nanoparticles (NPs) to form plasmonic gaps of just $1-2$nm~\cite{Mertens2013,Thacker2014,Benz2015,deNijs2015,Sigle2015}. A common example is the nanoparticle-on-mirror (NPoM) configuration (see Figure~\ref{fig:polygons}A), where a NP is assembled on a flat metallic surface separated by a molecular monolayer~\cite{FelixBenz2016,Emboras2016,Mertens2016,Chikkaraddy2018,Brassat2018}. Such systems show extreme optical  behaviour~\cite{Yashima2016,Kishida2022,Huang2019,Hoang2015,Akselrod2014,Sugimoto2018,Baumberg2019}, with many unprecedented applications~\cite{Yang2020,Zhang2017,Carnegie2018} such as light-matter stong coupling at room temperature~\cite{Chikkaraddy2016a}, 
enhanced exciton photoluminescence~\cite{Liu2016}, nonlinear vibrational pumping~\cite{Tomita2018}, sensing, mid-infrared upconversion detectors~\cite{Xomalis2021} and hot-electron emission~\cite{Li2017}.

In recent years, tremendous effort has been invested on fabricating and synthesising NPs of various shapes~\cite{Grzelczak2020,Liz-Marzan2004} that has fuelled advances in biological sensing~\cite{Jiang2018}, hot electron generation~\cite{Li2017}, enhance non-linear processes~\cite{Panoiu2018} but most importantly tracking and sensing of chemical and photocatalytic reactions~\cite{deNijs2019,Cortes2020}. 
Of particular interest is the dependence of photocatalytic reactions on the NP shape~\cite{Gregorio2020,Zhang2021}, where recent experimental results have demonstrated that the NP facet can determine the reaction selectivity and kinetics. This kinetic reaction control has been attributed to a combination of a thermal effect generated by the plasmon modes, the gold atom crystalline structure at the NP facet adjacent to the molecule and the field enhancement generated by the plasmon mode that enhances light-matter interactions, with increasing evidence that the latter is a significant contributor. 
However, these pathways are all linked together, with the crystalline nature of gold determining the NP shape, its facets' shapes and the plasmonic modes that govern both the thermal effects and the excitation of a molecule. Therefore it is nearly impossible to perform a systematic experimental study, while a theoretical study can provide valuable information and guidance. 
However, most theoretical studies have so far focused on idealised spherical (or truncated spherical) nanoparticles~\cite{Mertens2016,Kishida2022,Huang2019,Akselrod2014,Tserkezis2015,Lombardi2016,Hoang2016,Rose2014,Kongsuwan2020}, ignoring the significant impact of the NP shape and its facets have on light-matter interactions in such extreme nanocavities.

Here,  we fully characterize the electromagnetic behaviour of plasmonic nanocavities, formed by commonly occurring polyhedral nanoparticle shapes.
Recent experimental work~\cite{Benz2016} has revealed that due to the crystalline nature of gold, spherical NPs commonly adopt during their synthesis one of three polyhedral shapes: (i) cuboctahedron, (ii) rhombicuboctahedron and (iii) decahedron ---as shown in Figure~\ref{fig:polygons}B-D. 
Such polyhedral NPs form nanocavities that have a significant impact on molecules residing in them, due to both the enhanced light-matter interaction and the crystalline nature of the specific facet forming the nanocavity. 
We use a quasi-normal mode (QNM) analysis~\cite{Bai2013,Yan2018,Sauvan2022,Wu2021,Sauvan2013,QNMEig} to decompose the plasmonic modes of the nanocavities and show that the NP shape and the nanocavity symmetry dominate both their near- and far-field behaviour. 
Through a recombination technique, we obtain the total far-field emission profile for a molecule placed at specific positions within each nanocavity, which can be measured experimentally, and reveals how energy couples in and out of the system~\cite{Chikkaraddy2016a,Bedingfield2022}. 
Surprisingly, we find that the same NP shape gives very different results when assembled on a mirror with a different facet, even if the two facets' shape and size are identical.  
This work paves the way towards understanding and controlling at an unprecedented level light-matter interactions and photocatalytic reactions in extreme but realistic plasmonic environments.

\section{Quasi-Normal Mode Analysis of Plasmonic Nanocavities}
\label{sec:Eigen}

Plasmonic nanoantennas are open systems that both efficiently radiative to the far-field and are dissipate energy due to metallic Ohmic losses. As energy is not conserved during electromagnetic oscillations, they form non-Hermitian systems that cannot be characterized with the natural method of normal modes. Instead, a QNM description is required which yields complex eigenfrequecies~\cite{Sauvan2013}. In recent years, there have been significant advancements in the development of various quasi-normal analyses for such open and dissipative nanophotonic environments, with some notable examples being: the time-domain~\cite{Ge2014,Kristensen2014}, the pole-search~\cite{Bai2013,Powell2014,Zheng2014}, and the auxiliary-field eigenvalue method~\cite{Yan2018,Sauvan2013,Lalanne2018}, all summarized in a recent review~\cite{Sauvan2022}.
Here, we use the QNM methodology developed by Lalanne et al.~\cite{Yan2018,Sauvan2013,Lalanne2018}, where the auxiliary-fields are used to account for the dispersive behaviour of metals to linearise the eigenvalue problem.  
Due to the diverging nature of QNMs~\cite{Sauvan2013}, we surround the structure with perfectly matched layers (PMLs) that absorb all the energy emitted to the far-field~\cite{Yan2018}, and can be used to normalize the modes.
This method returns each QNM's eigenvector (i.e. electromagnetic near-field vectors) and their corresponding complex eigenvalues  (i.e. eigenfrequencies ): $\tilde{\omega}_i = \omega_i - i \kappa_i$, where the real part ($\omega_i$) is the spectral resonant frequency of each QNM and the imaginary part ($\kappa_i$) the mode linewidth describing the mode losses attributed to both the radiative and dissipative decay channels~\cite{Kristensen2015a}. 
The methodology is briefly outlined here, and discussed in more detail elsewhere~\cite{Yan2018,Sauvan2013,Lalanne2018}. 

In general, QNMs for the plasmonic system are found by solving the source-free Maxwell's equations for the electric $\tilde{\mathbf{E}}_i (\mathbf{r})$ and magnetic $\tilde{\mathbf{H}}_i (\mathbf{r})$ field vectors, while satisfying the Sommerfeld radiation condition for outgoing waves. Here, we consider metallic structures with their electric permittivity described by an $N$-pole Drude-Lorentz model:
\begin{equation}
\varepsilon (\omega) = \varepsilon_{\infty} \left( 1 + \sum_{k=1}^{N} \frac{\omega_{p,k}^2}{\omega_{0,k}^2 - \omega^2 - i\gamma_k \omega} \right) \label{eq:EPermN} \ ,
\end{equation}
where $\omega_{p,k}$, $\omega_{0,k}$ and $\gamma_k$ are the plasma frequency, resonant frequency and decay rate of the $k$-th Drude-Lorentz pole respectively, and $\varepsilon_\infty$ the asymptotic electric permittivity. The  dispersion of the electric permittivity introduces a non-linearity into the QNM eigenvalue problem. However, the problem can be linearised through the introduction of a pair of auxiliary fields~\cite{Yan2018}: 
\begin{equation}
\tilde{\mathbf{P}}_{i,k}(\mathbf{r})= \frac{\varepsilon_\infty \omega_{p,k}^2}{\omega_{0,k}^2-\tilde{\omega}_i^2-i\gamma_k\tilde{\omega}_i} \tilde{\mathbf{E}}_i (\mathbf{r})
\end{equation}
\begin{equation}
\tilde{\mathbf{J}}_{i,k}(\mathbf{r})= -i \tilde{\omega}_i \tilde{\mathbf{P}}_{i,k}(\mathbf{r}) \ ,
\end{equation}
where $\tilde{\mathbf{P}}_{i,k}$ and $\tilde{\mathbf{J}}_{i,k}$ are respectively the auxiliary polarization and current vectors of the $i^{th}$ QNM and $k^{th}$ Drude-Lorentz pole of the metal. For a two-pole ($N=2$) Drude-Lorentz model, the linearised eigenvalue problem to be solved is given by: 
\begin{equation}
\begin{bmatrix}
0&	-i\mu_0^{-1}\nabla\times&	0&	0&	0&	0 \\
i\varepsilon_0^{-1}\nabla\times&	0&	0&	-i\varepsilon_\infty^{-1}&	0&	-i\varepsilon_\infty^{-1} \\
0&	0&	0&	i&	0&	0\\
0&	i\omega_{p,1}^2\varepsilon_\infty&	-i\omega_{0,1}^2&	-i\gamma_{1}&	0&	0\\
0&	0&	0&	0&	0&	i\\
0&	i\omega_{p,2}^2\varepsilon_\infty&	0&	0&	-i\omega_{0,2}^2&	-i\gamma_{2}
\end{bmatrix}
\begin{bmatrix}
\tilde{\mathbf{H}}\\
\tilde{\mathbf{E}}\\
\tilde{\mathbf{P}}_1\\
\tilde{\mathbf{J}}_1\\
\tilde{\mathbf{P}}_2\\
\tilde{\mathbf{J}}_2
\end{bmatrix}=\tilde{\omega_i}\begin{bmatrix}
\tilde{\mathbf{H}}\\
\tilde{\mathbf{E}}\\
\tilde{\mathbf{P}}_1\\
\tilde{\mathbf{J}}_1\\
\tilde{\mathbf{P}}_2\\
\tilde{\mathbf{J}}_2
\end{bmatrix} \ , \label{eq:linearized_sys}
\end{equation}
where a full description can be found in~\cite{Yan2018}. Using the QNM solver `QNMEig'~\cite{QNMEig} with  COMSOL Multiphysics~\cite{COMSOL}, we perform finite-element numerical eigensolutions of Maxwell equations for this linearised system~\cite{Sauvan2013}, for various polyhedral NPs assembled on a flat gold mirror. It should be noted that the small gap between the NP and the mirror, in combination with the complicated morphology of the polyhedral NPs increase enormously the computational cost of these calculations. Additionally to represent realistic NPs, we introduce a curvature to the edges of each polyhedral nanostructure (discussed in more detail in the Supp. Info.), which further increases the computational demands.

\section{Polyhedron-on-Mirror Nanocavities}
\label{sec:NF}

\begin{figure}
\centering
\includegraphics[width=\linewidth]{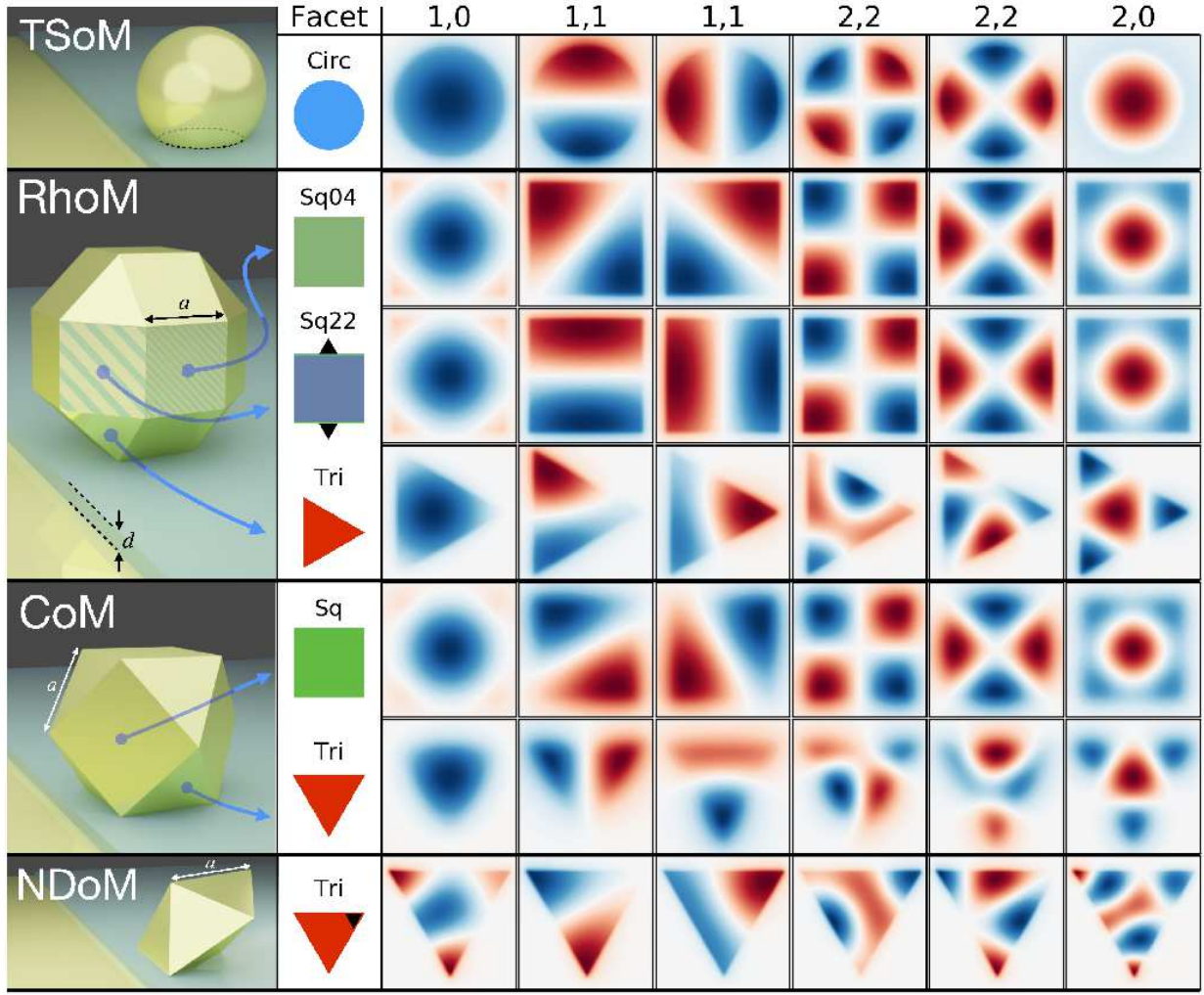}
\caption{\label{fig:near-field}QNMs of the different nanoparticle-on-mirror structures, with the facet forming the nanocavity indicated in the second column. The geometries from top to bottom are the: circular facet of TSoM; two square and one triangular facets of RhoM; singular square and triangular facets of the CoM; and the triangular facet of NDoM. The modes from left to right are: $(1,0)$, $(1,1)$, $(1,1)$, $(2,2)$, $(2,2)$, $(2,0)$, where the colour corresponds to the normalised QNM electric fields (Re[$E_{z,lm}$]) on the $xy$-plane through the centre of their respective nanocavities.}
\end{figure}

\begin{figure}[b!]
\centering
\includegraphics[width=\linewidth]{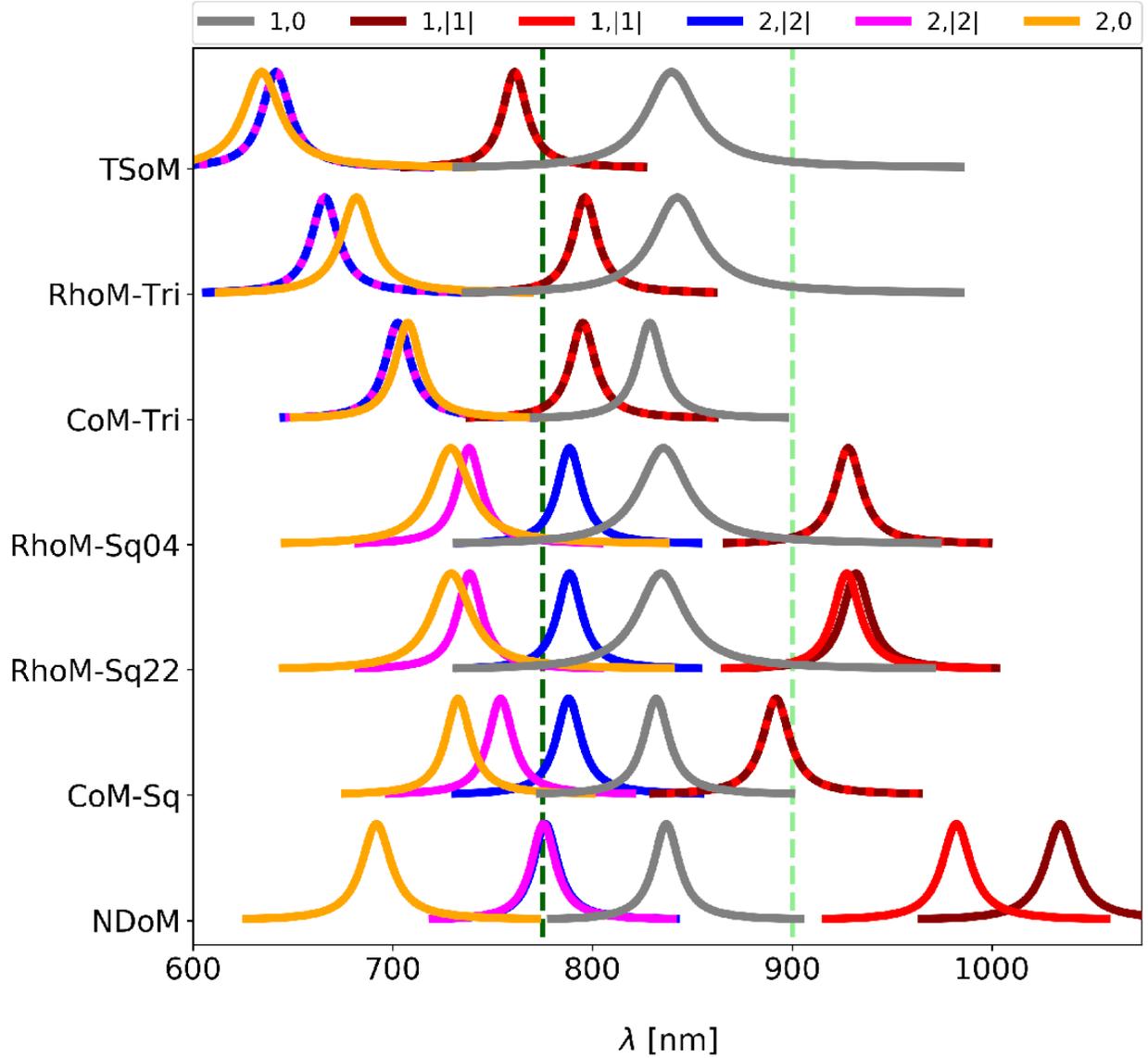}
\caption{\label{fig:radiative_loss} Complex QNM eigenfrequencies represented as Lorentzians, showing their spectral arrangement and energetic ordering for the: circular facet of the TSoM; two square and one triangular facets of the RhoM; singular square and triangular facets of the CoM; and the triangular facet of the NDoM. The labels correspond to the nanocavity configuration as shown in Figure~\ref{fig:near-field}. Dashed lines represent degenerate QNMs. Vertical dashed lines mark the wavelengths of (dark green) 775nm and (light green) 900nm. 
}
\end{figure}

During the synthesis of spherical NPs, the crystalline nature of the gold actually leads to the formation of polyhedral shapes: each exhibiting multiple facets of various shapes and different gold atom crystallization (see Figure~\ref{fig:polygons}) and when assembled on a mirror, tend to do so with one facet parallel to the mirror. So far, faceted NPoM geometries have been modelled theoretically mainly by truncating a perfectly spherical NP to form the Truncated Spherical NP on mirror (TSoM) geometry~\cite{Kongsuwan2020,Tserkezis2015}, which assembles on the mirror with a circular facet---as shown in Figure~\ref{fig:polygons}A. To allow comparison of previous studies with the polyhedral NPoM assemblies, we initially perform the QNM analysis for the TSoM geometry with: NP of radius $r_p = 40$nm, circular facet of diameter $f_d=20$nm, and spacer of refractive index $n=1.45$ and  thickness $d=1$nm. Figure~\ref{fig:near-field} (top row) shows the first six normalised electric field distributions on a plane through the centre of the nanocavity (and parallel to the mirror) in order of increasing energy, highlighting the clear mode confinement within the bounds of the circular facet. 
Although many methods that define the mode nomenclature have been introduced in the past\cite{Bozhevolnyi2007,Bozhevolnyi2008,Jung2009,Pors2013,Tserkezis2015}, here we use the spherical harmonic nomenclature introduced in~\cite{Kongsuwan2020} for identifying and labelling the set of supported QNMs, with each QNM labelled according to $i = (l,m)$---for a set of positive integers $l\geq 1$ and $-l \leq m \leq l$. The nomenclature shows that all modes with $m=0$ have an anti-node at the centre of the nanocavity, and all $m\neq0$ modes have a central node ~\cite{Kongsuwan2020}. 
Note that the cylindrical symmetry of the facet means that $(l,|m|)$-mode pairs are energetically degenerate, with their numerically identified orientation angle been arbitrary, but always orthogonal to each other. 
To aid the visualization of the modes' energetic ordering and degeneracies, we show in Figure~\ref{fig:radiative_loss} their spectral behaviour for TSoM (top row). Each QNM's complex eigenfrequency ($\tilde{\omega}$) obtained from the numerical QNM calculations is fitted to a Lorentzian (see Methods), such that its resonant frequency and linewidth are respectively described by $Re\{\tilde{\omega}\}$ and $Im\{\tilde{\omega}\}$. Note that dashed lines are used to represent degenerate modes. 

Here we consider the three most common NP shapes that occur during the synthesis of spherical gold NPs: (a) rhombicuboctahedron, (b) cuboctahedron, and (c) decahedron---shown in Figure~\ref{fig:polygons}, together with scanning electron microscopy images of spherical gold NPs, and the gold atom crystalization $\{100\}$, $\{110\}$ and $\{111\}$ shown in green, red and blue colours. 
When these polyhedral NPs are assembled on a mirror, they yield unique nanocavity systems due to the multiple distinct facets of each NP that can be assembled onto the mirror (see Figure~\ref{fig:near-field})~\cite{Elliott2022}. 
More specifically, the rhombicuboctahedron-on-mirror (RhoM) configuration has three unique NP arrangements on the mirror: two square facets, one with four square neighbouring facets (Sq04) and one with two square and two triangular neighbours (Sq22); and one triangular facet (Tri)~\cite{Bedingfield2022}. 
The cuboctahedron-on-mirror (CoM) system has two unique NP arrangements on the mirror---one square (Sq) and one triangular (Tri) facet---whereas the decahedron NP has ten identical triangular (Tri) facets, and therefore only offers a singular unique facet with which the nanodecahedron-on-mirror (NDoM) nanocavity can be assembled.
When a NP is assembled on a mirror, each of these facets forms a unique NPoM nanocavity.
Note that for each polyhedral shape, its edges are equal in length, so instead of a NP radius, these geometries can be parameterised by their characteristic edge length, which in this paper are chosen to be $a_\text{RhoM}=30$nm, $a_\text{CoM}=41$nm, and $a_\text{NDoM}=47$nm for the rhombicuboctahedron, cuboctahedron and decahedron NPs respectively. 
These values were chosen such that their respective $(1,0)$ nanocavity modes have similar resonant frequencies ($Re\{\tilde{\omega}\}$), which helps highlight the relative spectral positioning of all the other modes.
For each of these nanocavities, a background material of air is considered, along with the same spacer of thickness $d=1$nm and refractive index $n=1.45$. To ensure that our calculations represent realistic systems, we also rounded the edges of the polyhedral NPs with a curvature of $\rho=1$nm (see Figure~S2 for definition). This means that each facet edge is formed by approximately three gold atoms, which further increases the computational demands on our calculations. 

We start with the RhoM geometry assembled on its Sq04 facet (RhoM-Sq04)that is formed with $\{100\}$ gold atom crystallization, and has QNMs as shown in the second row of Figure~\ref{fig:near-field}. 
Note that the modes shown in Figure~\ref{fig:near-field} are not energetically ordered, but ordered according to the TSoM modes to aid in their comparison. 
Although the RhoM-Sq04 nanocavity does not have a cylindrical symmetry, it is apparent that the same spherical harmonic nomenclature can still be adopted to adequately identify and describe these modes. Here the $(1,0)$ mode is distorted to the square shape of the facet, but retains the same spatial characteristics with a circular central anti-node, as does the RhoM-Sq04 $(2,0)$ mode. However, unlike the arbitrary orientation of $(1,1)$-TSoM modes, the $(1,1)$ RhoM-Sq04 modes preferentially orientate towards the facet corners, due to the greater charge concentration there. 
Due to the four-fold symmetry of the RhoM when assembled on this facet, the $(1,1)$ modes remain orthogonal and are energetically degenerate. 
However, this four-fold nanocavity symmetry causes the two $(2,2)$ modes to lose their degeneracy (see Figure~\ref{fig:radiative_loss}). One of the $(2,2)$ modes has its anti-nodes in the corners of the facet---where the charges are energetically more favourably concentrated---while the second $(2,2)$ mode has to be orthogonal to the first $(2,2)$ mode and therefore is arranged with its anti-nodes at the less-confined facet edges. 

The Sq22 facet of the rhombicuboctahedron NP has $\{111\}$ gold atom crystallization, but the same size and shape as the Sq04 facet and therefore forms a geometrically identical nanocavity when assembled on the mirror (RhoM-Sq22). One can therefore easily assume that the electromagnetic QNMs in the gap of the RhoM-Sq22 nanocavity would be identical to those of RhoM-Sq04. However, the RhoM-Sq22 system carries a two-fold symmetry due to the pair of triangular facets neighbouring the Sq22 facet---as indicated by the small black triangles in the RhoM-Sq22 inset of Figure~\ref{fig:near-field}. 
These triangular facets lie at an inter-facet angle slightly closer to the mirror compared to the neighbouring square facets, which increases the field confinement along the edges adjacent to them. 
This makes the mode field appear `elongated' due to an unequal charge confinement supplied by the neighbouring square and triangular facets (see S.I. for more discussion). 
Although the $m=0$ modes are largely unchanged when compared to the RhoM-Sq04 modes, the broken geometrical symmetry of the RhoM-Sq22 nanocavity has a significant impact on the rest of the modes. 
For example, the pair of $(1,1)$ modes reorientate to now align with the sides of the facet rather than the corners---preserving their orthogonality. The unequal charge confinement across the facet breaks the degeneracy of the $(1,1)$ modes (see Figure~\ref{fig:radiative_loss}), with the mode directed between the neighbouring two triangular facets being more energetically favourable. 
Interestingly, the $(2,2)$ modes retain their orientations when compared to their RhoM-Sq04 counterparts, but due to the unequal charge confinement between the edges and corners of the Sq22 facet, the second $(2,2)$ mode loses its central node---with significant consequences to its far-field emission, as discussed later.

The final unique orientation of the RhoM geometry sees the assembly of the NP on its triangular facet (RhoM-Tri). Although the morphology of this facet is significantly different to both the circular and square facets seen so far, the same spherical harmonic nomenclature applies. The $(1,0)$ mode is clearly identifiable, with a single central anti-node confined to the bounds of the triangular facet. Due to the mismatch between the two lobes of the $(1,1)$ modes and the three-fold symmetry of RhoM-Tri system, the $(1,1)$ modes are less energetically favourable than the $(1,0)$ mode; however, they remain orthogonal to each other and are energetically degenerate (see Figure~\ref{fig:radiative_loss}). Similarly, the $(2,2)$ modes remain degenerate, but the three-fold symmetry of the geometry alters their charge distributions such that the nodal lines do not cross. The $(2,0)$ mode now exhibits stronger fields concentrations in the three corners of thee facet, but otherwise remains unchanged. Even though we have so far discussed the nanocavities formed only by the rhombicuboctahedron NP, the formation of three distinct nanocavities have been identified---each exhibiting a different optical response. 

We next consider the cuboctahedron NP assembled on the mirror, which can create nanocavities with either a square facet with $\{100\}$ gold atom crystallization or triangular facet with $\{110\}$ crystallization. When assembled on its square facet (CoM-Sq), it has the same facet shape and four-fold symmetry as the RhoM-Sq04, and supports an almost identical set of QNMs---with energetically degenerate $(1,1)$ modes and spectrally split $(2,2)$ modes. 
When instead assembled on the mirror with its triangular facet (CoM-Tri), the facet shape and three-fold symmetry of the CoM-Tri geometry leads to a very similar set of QNMs as those supported by the RhoM-Tri. 
Finally, we consider the decahedral NP assembled on a mirror (NDoM) that has  ten identical triangular facets with $\{110\}$ gold atom crystallization and allows for a single unique facet assembly on the mirror (see Figure~\ref{fig:near-field}). The slanting of the decahedral NP leads to a strong asymmetry in the field confinement across the facet, with the black triangle in the inset of Figure~\ref{fig:near-field} (bottom row) indicating the corner of the facet that sits underneath the NP centroid and receives the weakest confinement. 
This unequal field confinement leads to an `effective' facet centre that lies slightly closer to the corner under the NP, instead of the equilateral triangular facet's centre. This distorts the set of QNMs supported by the NDoM nanocavity, compared to those observed for the RhoM-Tri and CoM-Tri geometries. The $(1,0)$ mode is most affected by this asymmetry, which appears to push the charge distributions towards the corner under the NP---so much so that additional concentric anti-nodes appear. 
A similar behaviour is observed for higher order $m=0$ modes, with the $(2,0)$ mode changing profile significantly. 
Although the $(1,1)$ modes appear similar to those of the RhoM-Tri and CoM-Tri, the unequal confinement across the facet leads to the loss of their degeneracy and a large spectral separation (Figure~\ref{fig:radiative_loss}). Due to the form of the $(2,2)$ modes, however, they are negligibly affected by this weakly confined corner underneath the NP and are only very slightly non-degenerate.

Finally, Figure~\ref{fig:radiative_loss} shows the spectral position of the modes for all the nanocavity configurations discussed so far. We have chosen to set the size of the three NPs discussed here to have edge lengths of $a_{RhoM}=30$nm, $a_{CoM}=41$nm and $a_{NDoM}=47$nm, such that their $(1,0)$ mode to be resonant at similar wavelengths. This allows us to discuss the relative spectral position of the dark $(1,1)$ and $(2,2)$ modes with respect to their bright $(1,0)$ and $(2,0)$ modes, including their degeneracies. For example, the $(1,1)$ modes of RhoM-Sq22 nanocavity slightly shift compared to the RhoM-Sq04 and loose their degeneracy, since the system's four-fold symmetry has been broken (even though the facets forming the two nanocavities are identical in size and shape).
Now, comparing the two RhoM-square facet nanocavities with the RhoM-Tri, one sees large shifts for the $(1,1)$, $(2,0)$ and $(2,2)$ modes and a small red-shift for the $(1,0)$ mode, even though it is the same NP that forms all three nanocavities, assembled on the mirror with a different facet. 
One sees similar dramatic changes to the spectral position of modes for the cuboctahedron NP, when assembled on the mirror with either its triangular (CoM-Tri) or square (CoM-Sq) facet.
Hence the spectral position of the modes and the dark mode degeneracy strongly depends on the geometry of the NP and the symmetry of the overall system in general.

\section{Far-Field Emission Profiles}
\label{sec:FF}

Having obtained a complete characterization of the near-field and spectral behaviour of these polyhedron-on-mirror nanocavities, we now demonstrate how each nanocavity out-couples energy to the far-field, and therefore how a molecule can transfer energy to the far-field to be measured experimentally. Using reciprocity, we perform a Near-to-Far-Field Transformation (NFFT) for the QNM near-fields shown in Figure~\ref{fig:near-field}, and obtain their far-field Poynting flux with the software RETOP~\cite{RETOPSoftware,RETOP,Kongsuwan2020,Pors2015,Balanis2016,Demarest1996}. 
This method first considers the QNM electric $\tilde{\mathbf{E}}_{lm} (\mathbf{r})$ and magnetic $\tilde{\mathbf{H}}_{lm} (\mathbf{r})$ fields of each $(l,m)$ mode on a region enclosing the NP, ensuring that it intersects all material layers of the system: in our case this is the mirror, spacer and air domains. 
These near-fields $(\tilde{\mathbf{E}}_{lm} (\mathbf{r}), \tilde{\mathbf{H}}_{lm} (\mathbf{r}))$ are then projected to the far-field ($\tilde{\mathbf{E}}^{ff}_{lm}, \tilde{\mathbf{H}}^{ff}_{lm}$)$e^{i \omega_{em} R/c}$ on a hemispherical dome of radius $R=1$m above the NPoM geometry at the molecule frequency $\omega_{em}$, from which the time-average Poynting flux $\langle S_{lm} \rangle = \text{Re}[\tilde{\mathbf{E}}^{ff \ *}_{lm} \times \tilde{\mathbf{H}}^{ff}_{lm}]/2$ is obtained for each $(l,m)$ mode as a function of the polar ($\theta$) and azimuthal ($\phi$) angles of the dome (where the polar angle is zero at the top of the dome). 

\begin{figure}[b!]
\centering
\includegraphics[width=\linewidth]{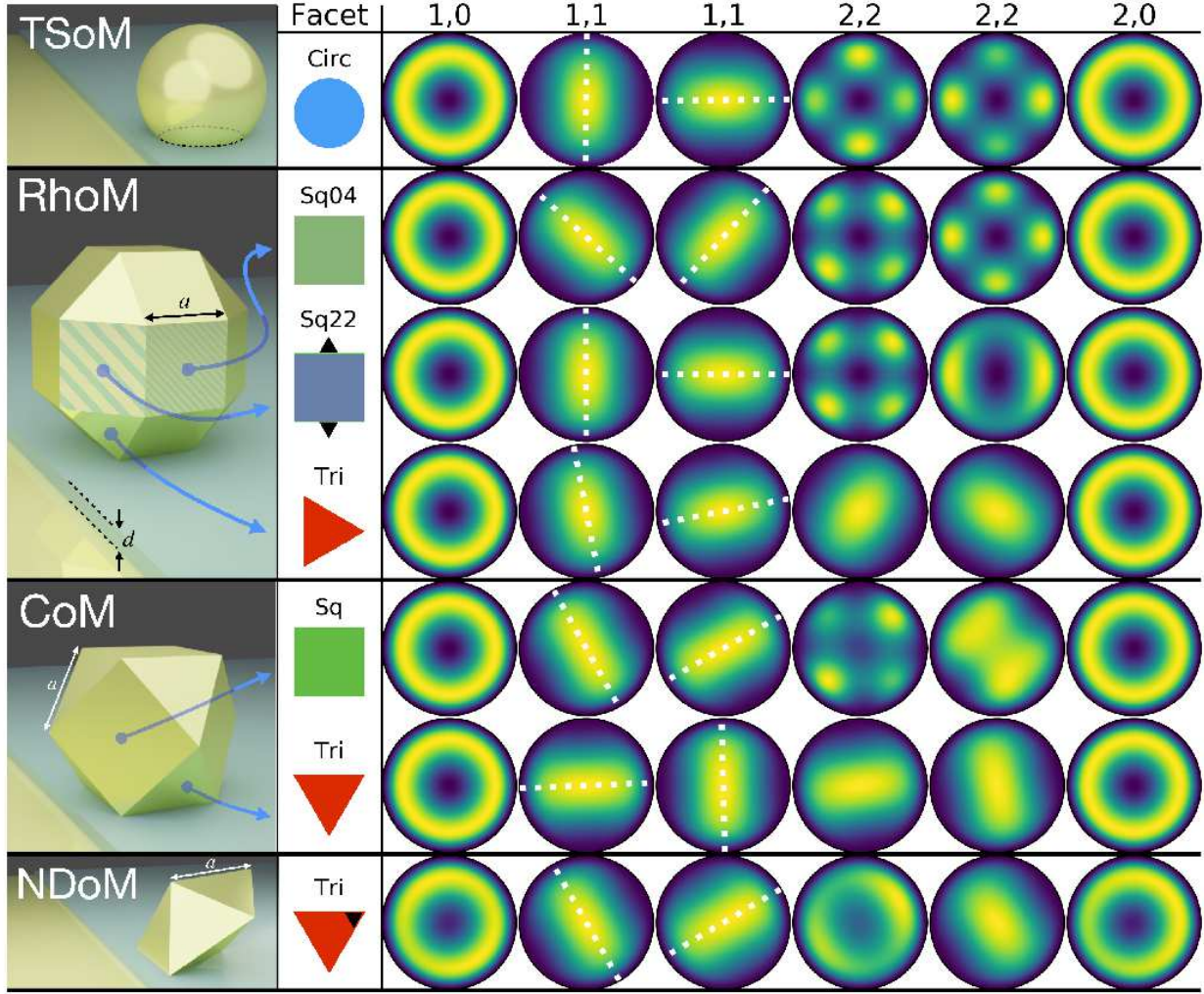}
\caption{\label{fig:far-field}Far-field emission for the QNMs of the different particle-on-mirror structures, with the facet forming the nanocavity indicated in the second column. The geometries from top to bottom: circular facet of the TSoM; two square and one triangular facets of the RhoM; singular square and triangular facets of the CoM; and the triangular facet of the NDoM. The modes from left to right: $(1,0)$, $(1,1)$, $(1,1)$, $(2,2)$, $(2,2)$, $(2,0)$, where the colour corresponds to the normalised time averaged Poynting flux $\langle S_{lm} \rangle$. White dashed lines are added over the $(1,1)$ modes to highlight the correspondence of their direction and orthogonality with their near-field counterparts (i.e. normal to each other).}
\end{figure}

The far-field emission patterns for all the modes and nanocavities discussed previously are shown in Figure~\ref{fig:far-field}. 
For the TSoM geometry with its cylindrically symmetric circular facet~\cite{Kongsuwan2020}, all $m=0$ modes emit in a conical shape with the maximum energy at $\theta = 62^\circ$~\cite{Baumberg2019,Chikkaraddy2017,Mubeen2012}, whereas the pair of $(1,1)$ modes largely emit normally away from the mirror with smaller contributions following their near-field lobes (as shown by the white dashed lines). 
All other $m>1$ modes exhibit $2m$-lobes of equal intensity, emitted conically away from the mirror at the same angle of $\theta = 62^\circ$, with each pair of $m\neq0$ modes retaining their mode orthogonality to the far-field. 
It should be noted that the NFFT used to obtain the far-field emissions in Figure~\ref{fig:far-field} is extremely sensitive to numerical errors in the QNM electromagnetic fields $(\tilde{\mathbf{E}}_{lm} (\mathbf{r}),\tilde{\mathbf{H}}_{lm} (\mathbf{r}))$ ---with increasingly finer meshing required for higher order modes that have finer near-field features.
For this reason, the $(3,3)$ TSoM mode does not emit normally away from the mirror as it was previously reports in~\cite{Kongsuwan2020}, but in fact has six conically emitting lobes (see Supp. Info.). 
The different lobe intensities are also due to numerical errors and it should be noted that it is computationally too expensive to distinguish the far-field lobes for modes of $m>3$.

Returning to the polyhedron-on-mirror nanocavities, the far-field emission patterns are shown in Figure~\ref{fig:far-field}. All $m=0$ modes retain their conical emission shape at $\theta = 62^\circ$, regardless of the nanocavity geometry or NP symmetry. 
The $(1,1)$ modes are similarly consistent for all nanocavities, emitting largely normal from the mirror, with only their azimuthal orientation varying to follow the near-field lobe orientation and therefore preserve the mode orthogonality. 
The biggest differences are seen for the $(2,2)$ modes. Most nanocavities formed with a square facet (RhoM-Sq44, RhoM-Sq22, CoM-Sq) retain the four-lobed far-field emission of the $(2,2)$ modes. 
A notable exception is the second $(2,2)$ mode of the RhoM-Sq22 nanocavity, whose near-field appears as if `elongated' due to the more shallow inclination of the neighbouring triangular facets compared to the neighbouring square facets (see Suppl. Info. for more discussion)---which breaks the four-fold symmetry of the facet. 
As discussed earlier, this effective mode `elongation' destroys the central near-field node, with not $E_{z}\neq 0 $ for this $(2,2)$ mode at the centre of the nanocavity. 
This has the remarkable ability to completely dominate the emissive response of the system that now acquires a more conical far-field emission---with the peak intensity again at the same $\theta = 62^\circ$.

For nanocavities formed with a triangular facet (RhoM-Tri, CoM-Tri, NDoM), again the $m=0$ and $(1,1)$ modes retain the same far-field emissions patterns as the TSoM nanocavity, but the $(2,2)$ modes lose their four-lobed emission pattern. Interestingly, they now emit normally away from the mirror, and in a very similar way to $(1,1)$ modes---with almost identical far-field patterns observed for both the RhoM-Tri and CoM-Tri nanocavities. 
This is due to the destroyed four-fold symmetry of the $(2,2)$ modes in the near-field, since they are `squeezed' within the triangular nanocavities.
So now, the `overall' far-field response is dominated by the dipoles that form between the modes' anti-nodes at the near-field nanocavity charge distributions. 
A particular case is the NDoM nanocavity, where the first $(2,2)$ mode shows conical emission to the far-field. This is due to the asymmetric geometry of the system that causes unequal confinement across the triangular facet. This effectively shifting the centre of the facet (indicated with a black triangle in Figure~\ref{fig:far-field}). 
It also loses its central lobe in the near-field; and therefore its far-field emission is a superposition of a conical emission (similar to $m=0$ modes) and a normal emission (similar to a $(1,1)$ mode). 
The second $(2,2)$ mode now resembles the $(1,1)$ mode in the near-field, and it therefore emits to the far-field in a similar manner---normally away from the mirror. 
Finally, this asymetric geometry of the system also impacts the far-field emission of the $(2,0)$ mode which, although still conical, is now asymmetric---with higher intensity along the azimuthal angle that aligns with the `effective' centre of the triangular facet (indicated with a black triangle in Figure~\ref{fig:far-field}).

\begin{figure}
\centering
\includegraphics[width=0.75\linewidth]{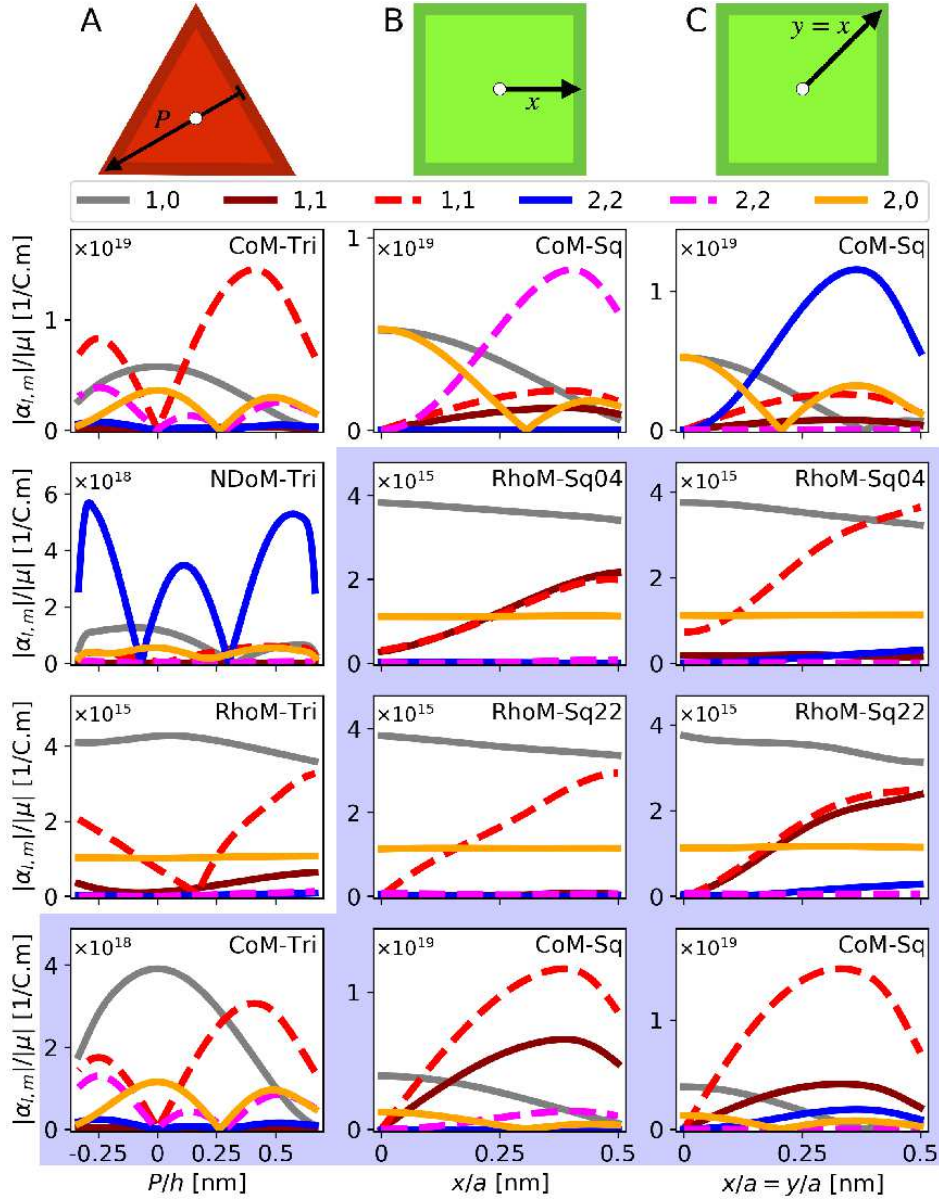}
\caption{\label{fig:alphas}$\alpha$-coefficients of the polyhedral NPoM geometries, for a series of emitter positions within the nanocavities. Column A - The triangular facets of the polyhedral systems, following a path from perpendicular line drawn from one of the facet edges to the opposite corner, through the triangular facet centroid---normalised to the total length of this parth ($h=\sqrt{3}a/2$). Column B - The square facets of the polyhedral systems, following a path along the $x$-axis from the centre of the facet. Column C - The square facets of the polyhedral systems, following a path along the diagonal from the centre of the facet. White and blue backgrounds respectively correspond to emitter transition wavelengths of $\lambda_{em}=775$nm and $\lambda_{em}=900$nm.}
\end{figure}

Although the far-field profile of each individual mode is very revealing of each mode's behaviour, when a molecule resides in such nanocavities actually couples to many modes simultaneously, 
forming an overall optical response that is dependent on the superposition of both the spatial and spectral properties of the QNM modes and the emitter. 
This collective response can be obtained using a recombination technique and utilizing the electromagnetic reciprocity of the system~\cite{Kongsuwan2020,Sauvan2013} as:
\begin{equation}
\{ \tilde{\mathbf{E}}(\mathbf{r}), \tilde{\mathbf{H}}(\mathbf{r}) \} = \sum_{lm} \alpha_{lm} (\mathbf{r}_{em}, \omega_{em}) \{ \tilde{\mathbf{E}}_{lm} (\mathbf{r}), \tilde{\mathbf{H}}_{lm} (\mathbf{r}) \} \ ,
\end{equation}
where $\{ \tilde{\mathbf{E}}_{lm} (\mathbf{r}), \tilde{\mathbf{H}}_{lm} (\mathbf{r}) \}$ are the electromagnetic near-fields of mode $(l,m)$ and $\{ \tilde{\mathbf{E}}(\mathbf{r}), \tilde{\mathbf{H}}(\mathbf{r}) \}$ are the total fields. The excitation coefficient of mode $(l,m)$ is $\alpha_{lm} (\mathbf{r}_{em}, \omega_{em})$  and is dependent on both the position of the emitter $\mathbf{r}_{em}$ and its resonant frequency $\omega_{em}$, given by~\cite{Sauvan2013}: 
\begin{align}
\alpha_{lm} (\mathbf{r}_{em}, \omega_{em}) = - \omega \sum_{l'm'} B^{-1}_{lm,l'm'} (\omega_{em}) \mathbf{\mu}_{em} \cdot \tilde{\mathbf{E}}_{l'm'} (\mathbf{r}_{em}) \ ,
\end{align}
where $\mathbf{\mu}_{em}$ is the dipole moment of the emitter, and the term $B_{lm,l'm'} (\omega)$ is a matrix constructed as~\cite{Kongsuwan2020,Sauvan2013}: 
\begin{align}
B_{lm,l'm'} (\omega) = \iiint_{\Omega} \tilde{\mathbf{E}}_{l'm'} \cdot \left[ \omega \varepsilon (\mathbf{r}, \omega) - \tilde{\omega}_{lm} \varepsilon (\mathbf{r}, \tilde{\omega}_{lm}) \right] \tilde{\mathbf{E}}_{lm} - \mu_0 \tilde{\mathbf{H}}_{l'm'} \cdot \left( \omega - \tilde{\omega}_{lm} \right) \tilde{\mathbf{H}}_{lm} \ d\mathbf{r}^3 \ ,
\end{align}
for which we included twenty modes for each nanocavity considered here. 
This is in agreement with previous results~\cite{Kongsuwan2020} where it was shown that twenty modes are an adequate number to obtain the overall behaviour. 
The total far-field emission of each structure is related to its modal components, using the same set of $\alpha$-coefficients, as~\cite{Kongsuwan2020,Sauvan2013}: 
\begin{equation}
\{ \tilde{\mathbf{E}}^{ff} (\mathbf{r}), \tilde{\mathbf{H}}^{ff} (\mathbf{r}) \} = \sum_{lm} \alpha_{lm} (\mathbf{r}_{em}, \omega_{em}) \{ \tilde{\mathbf{E}}^{ff}_{lm} (\mathbf{r}), \tilde{\mathbf{H}}^{ff}_{lm} (\mathbf{r}) \} \ . 
\end{equation}
The total time-average Poynting flux in the far-field is therefore given by $\langle S_{tot} \rangle = \text{Re}[\tilde{\mathbf{E}}^{ff \ *} \times \tilde{\mathbf{H}}^{ff}]/2$, and is dependent on both the transition frequency $\omega_{em}$ of the emitter (or its corresponding wavelength $\lambda_{em}$), and its position within the nanocavity $\mathbf{r}_{em}$~\cite{Kongsuwan2020}. 
The resonant frequency of the emitter $\omega_{em}$ determines which of the nanocavity's modes are excited. 
It should be noted that the spectral behaviour and local density of states of these nanocavities can be constructed from the $\alpha_{lm} (\mathbf{r}_{em}, \omega_{em})$ values, and has been shown in~\cite{Elliott2022}.
By consulting the spectral arrangement of each nanocavity's modes (as shown in Figure~\ref{fig:radiative_loss}), we identify two interesting frequencies for a dipole emitter, $\lambda_{em}=775$nm and $\lambda_{em}=900$nm, one at either side of the $(1,0)$ mode, shown as dark and light green vertical dashed lines. 
These emitter frequencies are chosen to highlight how the different spectral position of the dark modes ($(1,1), (2,2)$ etc) with respect to the primary bright mode $(1,0)$, determine how a molecule couples energy out of the system and how this is manifested in a far-field emission pattern that can be measured experimentally. 
Although in general one would expect that very off-resonant modes would not couple with the molecule and have negligible $\alpha$-coefficient values, this is not always the case. For example, the $(1,1)$ modes of the CoM-Sq nanocavity and a molecule with $\lambda_{em}=775$nm (Figure~\ref{fig:alphas} top row, second and third column), where the modes are resonant at $\lambda_{(1,1)}=1100$nm, but are still excited, producing $\alpha$-coefficient values comparable to the more resonant mode $(2,0)$. This is due to the field enhancement of the $(1,1)$ modes that is significantly larger than other modes. 
%the $\alpha$-coefficients and therefore the far-field emission changes as different molecules are placed within the nanocavities, as well as how the position of the emitter changes the behaviour. 

We start with the triangular-facet nanocavities (i.e. CoM-Tri, RhoM-Tri, NDoM-Tri), for a molecule placed at positions $P$ across the height  $h$  of the equilateral triangular facet that passes through its centroid (or ortho-centre), as shown atop Figure~\ref{fig:alphas} Column A, and plot the corresponding $\alpha$-coefficients.
To ensure consistency between our results for each system, the orientation of the path is chosen to follow the nodal line of the first $(1,1)$ mode's near-field. 
Note that $P$ is measured relative to the centre of the equilateral triangular facet, and that we only show here the interesting results, with figure with white (blue) backgrounds show the $\alpha$-coefficients for a dipole emitter resonant at $\lambda_{em}=775$nm ($\lambda_{em}=900$nm). 
For completeness, we also present all other results in Supp. Info.  
High $\alpha$-coefficient values reveal the modes with which molecules of $\lambda_{em}=775$nm and $\lambda_{em}=900$nm primarily couple to, and with which the system out-couples energy to the far-field. 
%Although in general one would expect that very off-resonant modes would not couple with the molecule, this is not always the case. For example, the $(1,1)$ modes of CoM-Sq and a molecule with $\lambda_{em}=775$nm (Figure~\ref{fig:alphas} top row, second and third column), where the modes are resonant at $\lambda=1100$nm, but are still excited and contribute tot he far-field emission. This is due to the field enhancement of the $(1,1)$ modes that is significantly larger than for the two $(2,2)$ modes that the molecule is spectrally more close. 

\begin{figure}[t!]
\centering
\includegraphics[width=\linewidth]{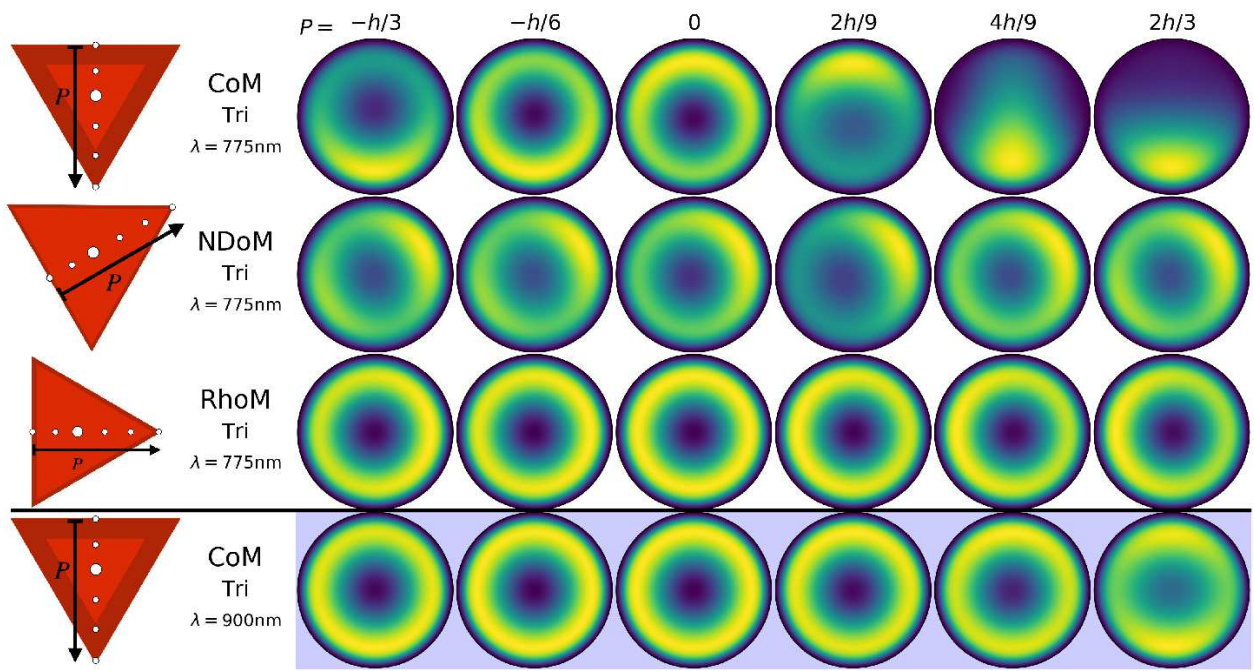}
\caption{\label{fig:SrReTri}Reconstructed total far-field emission of the different particle-on-mirror structures, for a series of emitter positions within the nanocavities. The geometries are the CoM, NDoM and RhoM structures assembled on their triangular facets, with the colour corresponding to the normalised time averaged total Poynting flux $\langle S_{tot} \rangle$. The emitter positions follow a path $P$ from the perpendicular line drawn from one of the facet edges to the opposite corner of the triangle facet, through its centroid---and is shown normalised to the length of this path ($h=\sqrt{3}a/2$). White and blue backgrounds respectively correspond to emitter transition wavelengths of $\lambda_{em}=775$nm and $\lambda_{em}=900$nm, and the dark red border represents the relative rounding region of the facet edge with respect to the facet size (i.e. $\rho/a$) of each nanocavity.}
\end{figure}

In Figure~\ref{fig:SrReTri} we show how the overall far-field emission pattern changes with the spatial position of the molecule. The position of a $\lambda_{em}=775$nm molecule within the RhoM-Tri nanocavity does not significantly affect the emission pattern, with only minor differences observed when the molecule is at the tip of the triangular facet, due to the stronger $(1,1)$ mode field enhancement that is almost resonant with the molecule. The conical shape of the RhoM-Tri emission pattern demonstrates that most of the molecule's energy is coupled into the $(1,0)$ mode, which is in line with the $\alpha$-coefficients shown in Figure~\ref{fig:alphas} (first column, third sub-figure). The NDoM-Tri nanocavity exhibits a `horse-shoe' like conical emission pattern for a $\lambda_{em}=775$nm molecules, but now this is primarily due to the first $(2,2)$ mode with which the emitter is exactly on resonance with, and with some added contribution from both the $(1,0)$. $(2,0)$ and $(1,1)$ modes. 
As the emitter molecule moves across the triangular facet, it passes through the first $(2,2)$ mode's node (at $P=2h/9$), which maximizes one side of the conical emission. This asymmetric emission pattern is due to the combined excitation of $(1,0)$, $(2,0)$ and $(1,1)$ modes, which are the only ones with non-zero $\alpha$-coefficients at $P=2h/9$.
By placing the molecule further to the corner of the triangular nanocavity, it reverts back to the $(2,2)$ emission pattern, in line with the $\alpha$-coefficient results shown in Figure~\ref{fig:alphas} (first column, second figure), where the $(2,2)$ mode dominates for a $\lambda_{em}=775$nm molecule. 

It is more interesting though to look at the CoM-Tri nanocavity and compare its emission patterns in Figure~\ref{fig:SrReTri} for molecules with either $\lambda_{em}=775$nm or $\lambda_{em}=900$nm, which are distinctively different. 
The two molecule wavelengths were chosen to be at either side of the $(1,0)$ mode, with the $\lambda_{em}=900$nm  predominantly coupling to both $(1,1)$ modes, with some small contribution from the $(1,0)$ mode. 
Therefore, it produces emission patterns with a single lobe, with its azimuthal position dependent on which $(1,1)$ mode has the dominant field enhancement at the position of the molecule, and is in agreement with the $\alpha$-coefficients shown in Figure~\ref{fig:alphas} (first column, last figure).
The $\lambda_{em}=775$nm molecule couples almost equally with the $(1,0)$, $(2,0)$ and one of the $(2,2)$ modes, but also couples to both $(1,1)$ modes even though they are spectrally away (i.e. $\lambda_{11}\sim 910$nm). The two $m=0$ modes produce conical emission patterns, with an exception at $P=2h/9$. At this position, the values of the $\alpha$-coefficients (see Figure~\ref{fig:alphas}, first column, top figure) show that the molecule couples almost equally to the $(1,0)$ and the two $(1,1)$ modes, while the $(2,0)$ and $(2,2)$ modes have a node. Therefore, producing an emission pattern of a single lobe only when the molecule is placed at $P=2h/9$.

%In contrast, the NDoM (that only forms triangular nanocavities) shows a very intriguing behaviour. The $\alpha$-coefficients for a molecule with $\lambda_{em}=775$nm are dominated by the first $(2,2)$ mode with some contribution from the $m=0$ modes, and a negligibly small $(1,1)$ contribution (see Figure~\ref{fig:alphas}). 
%This is also reflected in the far-field emission patterns (see Figure~\ref{fig:SrReTri}, second row), where the first $(2,2)$ mode primarily contributes. 
%When the emitter is positioned at $P=2h/9$, the $\alpha$-coefficient of the first $(2,2)$ mode becomes comparable in value with the second $(1,1)$ mode as well as the $m=0$ modes, which increases the asymmetry in the otherwise conical far-field emission (see Figure~\ref{fig:SrReTri}). 
%However, the first $(2,2)$ mode dominates most of the far-field emission, regardless of the molecule's position, and always exhibiting a `horse-shoe'-like emission pattern.
%It should be noted that the asymmetry of the decahedron NP induces an `effective' facet centre that is different to the geometrical centre ($P/h = 0$) of the triangular facet (as discussed earlier), which is clearly evident from the $\alpha$-coefficient of the $(1,0)$ mode that peaks at $P/h \sim 0.12$. 

\begin{figure}[b!]
\centering
\includegraphics[width=\linewidth]{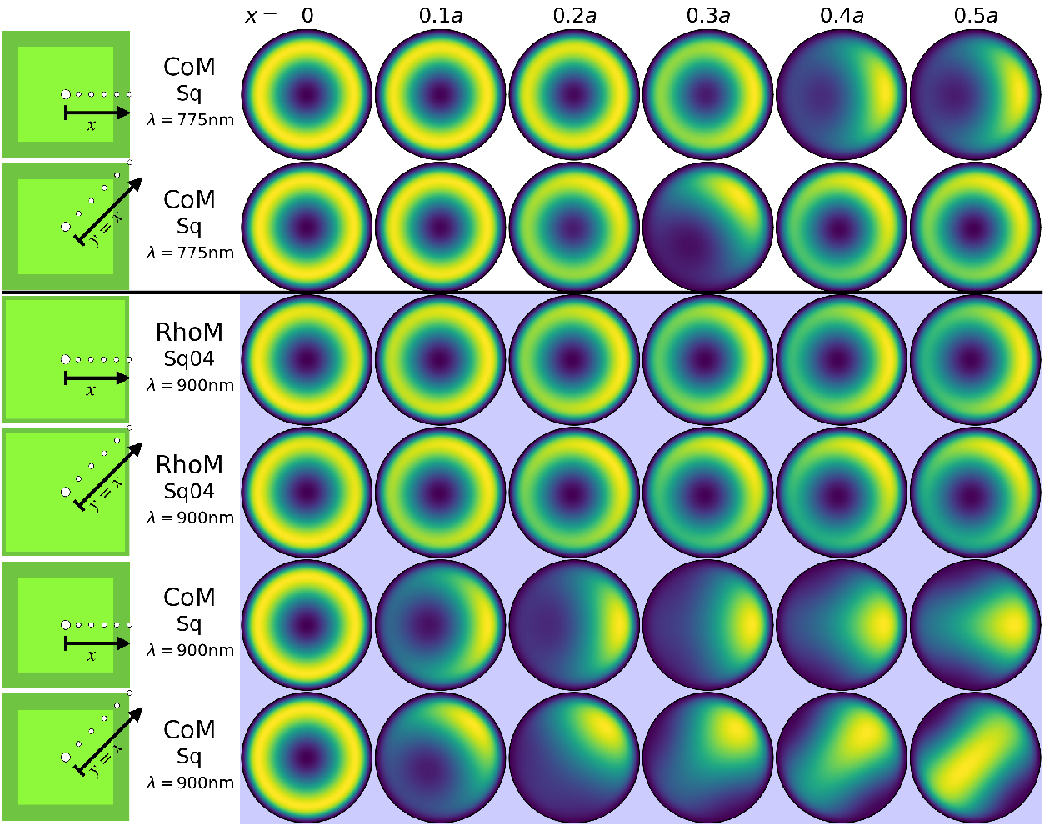}
\caption{\label{fig:SrReSq}Reconstructed total far-field emission of the different particle-on-mirror structures, for a series of emitter positions within the nanocavities. The geometries are the CoM and RhoM structures assembled on their square facets, with the colour corresponding to the normalised time averaged total Poynting flux $\langle S_{tot} \rangle$. For each structure, two emitter paths are considered: one from the centre of the facet to the edge, along the $x$-axis; and the other from the centre of the facet to the corner, along the diagonal. White and blue backgrounds respectively correspond to emitter transition wavelengths of $\lambda_{em}=775$nm and $\lambda_{em}=900$nm.}
\end{figure}

Similarly for square-facet nanocavities (i.e. CoM-Sq, RhoM-Sq04, RhoM-Sq22), for which we consider two molecule paths within the nanocavity. Both start at the centre of the facet, with one path tracing horizontally towards the facet edge (i.e. along the x-axis)---see Figure~\ref{fig:alphas}, Column B---and the other diagonally to the facet corner, see Figure~\ref{fig:alphas}, Column C for the $\alpha$ coefficients. 
We start with the CoM-Sq nanocavity and a $\lambda_{em}=775$nm molecule. In Figure~\ref{fig:SrReSq}, we see that for a molecule at the centre of the nanocavity the $(1,0)$ and $(2,0)$ modes are dominant, producing a conical emission pattern---but as the molecule moves towards the facet edge (corner), the second (first) $(2,2)$ mode starts to become more significant, interfering with the $(1,0)$ and $(2,0)$ modes to gradually produce a singular lobe emission pattern with its azimuthal position depending on the location of the molecule within the nanocavity.  
%This is also evident from the far-field emissions shown in Figure~\ref{fig:SrReSq} (first and second rows), where for a molecule at the centre of the facet, a conical emission is observed (coming from the $m=0$ modes), but as the emitter approaches the facet edge or corner, one of the $(2,2)$ modes dominates the far-field emission. 
Now, a $\lambda_{em}=900$nm molecule within the same nanocavity (CoM-Sq) produces very different emission patterns (see Figure~\ref{fig:SrReSq}).  
At this wavelength, the molecule couples very efficiently into the $(1,1)$ modes, with some contribution also from the $(1,0)$mode, in agreement with the $\alpha$-coefficients shown in Figure~\ref{fig:alphas} (second and third column, bottom figure). 
At the centre of the nanocavity, the system emits conically due to the $(1,0)$ mode, but as moving away from the centre, the $(1,1)$ modes dominate the far-field emission (Figure~\ref{fig:SrReSq}, bottom two rows).
Therefore, the molecule's emission wavelength determines which $m\neq 0$ mode out-couples the energy from the near- to the far-field.
Finally, the RhoM-Sq04 nanocavity with a $\lambda_{em}=900$nm molecule is considered, again moving from the centre of the square facet along either horizontal or diagonal paths. 
The $(1,0)$ mode dominates the emission pattern with a perfectly conical pattern, while the contributions to the far-field of both $(1,1)$ modes increases as the emitter moves horizontally, to produce again an asymmetric emission pattern. 
However, only the second $(1,1)$ mode contributes to the emission pattern for moving the molecule diagonally. Both results are in agreement with the $\alpha$-coefficient values shown in  Figure~\ref{fig:alphas} (second and third column, second figure). 
%This behaviour is also reflected in the far-field emission patterns shown in Figure~\ref{fig:SrReSq} (third and fourth rows), where an emitter at the centre of the square facet produces a perfect conical emission, but an asymmetric emission emerges as the emitter moves away from the centre. 
Note that for both cases the azimuthal placement of the far-field emission lobe follows the position of the molecule within the nanocavity. 

Hence, the polyhedral shapes that spherical gold NPs acquire during their synthesis dominate how light interact with molecules in realistic plasmonic nanocavities and most importantly how energy out-couples from the system to be measured experimentally.
Although the electromagnetic modes remain relatively robust, they significantly change spectrally for each design (see Figure~\ref{fig:radiative_loss}), and dominate the far-field emission patterns. The emission patterns that can be measured experimentally, contain significant information about the molecule's position within the nanocavity and about the nanocavity. Therefore it is very important to understand and characterize the electromagnetic behaviour of realistic nanocavities, especially if one wants to perform photocatalytic reactions or other light-matter interactions in plasmonic nanocavities, such as non-linear vibrational pumping. 
The work presented here is of particular importance for photocatalytic reactions of molecules in nanocavities, where the reaction dynamics (and manipulation of the molecule) is determined by the gold atom crystallization that forms the facet and shapes the NP, the thermal effect generated by the specific electromagnetic mode excited and its field enhancement. 
Since here we are considering realistic NP-shapes formed by the crystalline nature of gold, one can use our results together with experimental measurements to determine the impact of the gold crystallization for such photocatalytic reactions. 
So, our work paves the way towards understanding and controlling molecules and their photochemical reactions with such extreme light-matter interactions, by simply recording the system's far-field emission pattern. 

\section{Conclusion}
\label{sec:Conclusion}

Spherical gold nanoparticles acquire various polyhedral shapes during their synthesis due to the crystalline nature of gold. 
Here, using a quasi-normal mode analysis for nanocavities formed by the three most commonly occurring polyhedral nanoparticles (i.e. cuboctahedron, rhombicuboctahedron, decahedron) assembled on a flat metal surface, we show the rich multi-modal nature of their optical responses, that is dominated by the geometrical morphology of the nanogap and its neighbouring facets, and the symmetry of the overall system. 
We also provide the far-field emission from each nanocavity that shows the morphological-dependent intricacies with which the system is able to out-couple energy, as well as its dependence on the molecule's emission frequency and position within the nanocavity. 
Our results provide information on how molecules interact with plasmonic modes in realistic nanocavities, and allow for the interpretation of experimental measurements to determine the impact of gold atom crystallization for photocatalytic reaction.

\section{Methods}
\label{sec:Methods}

\subsection*{Optical Properties of Gold}
For each of the structures and facet orientations considered here, the QNM simulations see the polyhedral NP and spacer placed at the bottom of a cylindrical domain with a height and radius of $250$nm. This cylinder has $150$nm thick PMLs placed on top and around the sides, with a $100$nm thick substrate set beneath it. The gold NP and substrate in each system are modeled by a $2$-pole Drude-Lorentz electric permittivity. This reduces the general expression of equation~\ref{eq:EPermN} to: 
\begin{equation}
\varepsilon (\omega) = \varepsilon_\infty \left( 1 + \frac{\omega_{p,1}^2}{\omega_{0,1}^2 - \omega^2 - i\gamma_1 \omega} + \frac{\omega_{p,2}^2}{\omega_{0,2}^2 - \omega^2 - i\gamma_2 \omega} \right) \ ,
\end{equation}
where $\varepsilon_\infty = 6 \varepsilon_0$, $\omega_{p,1} = 5.37 \times 10^{15}$ rad/s, $\omega_{0,1} = 0$ rad/s, $\gamma_1 = 6.216 \times 10^{13}$ rad/s, $\omega_{p,2} = 2.2636 \times 10^{15}$ rad/s, $\omega_{0,2} = 4.572 \times 10^{15}$ rad/s, and $\gamma_2 = 1.332 \times 10^{15}$ rad/s. This is indirectly implemented into the numerical model via the pair of auxiliary fields for each pole. 

\subsection*{Lorentzian Fits to QNMs}
For the spectral representation of each system's modal distribution in Figure~\ref{fig:radiative_loss}, the real ($\omega_i$) and imaginary ($\kappa_i$) components of each QNM's eigenfrequency are fitted to a Lorentzian, as follows: 
\begin{equation}
L(\omega) = \frac{(\kappa_i/2)^2}{(\kappa_i/2)^2 + (\omega - \omega_i)^2} \ ,
\end{equation}
where $\kappa_i$ corresponds to the full-width half-maximum.

\medskip
\textbf{Supporting Information} \par 
Supporting Information is available.

% Acknowledgements
\medskip
\textbf{Acknowledgements} \par 
AD would like to acknowledge funding support from the Royal Society University Research Fellowship URF\textbackslash R1\textbackslash 180097, Royal Society Research Fellows Enhancement Award RGF \textbackslash EA\textbackslash 181038, Royal Society Research grants RGS \textbackslash R1\textbackslash 211093 and funding from EPSRC for the CDT in Topological Design EP/S02297X/1, EPSRC programme grant EP/X012689/1.
J.J.B. acknowledges EPSRC Grants EP/N016920/1, EP/L027151/1, NanoDTC EP/L015978/1, and support from European Research Council (ERC) under Horizon 2020 research and innovation programme THOR 829067, POSEIDON 861950 and PICOFORCE 883703.
\medskip

\bibliographystyle{unsrt}
\bibliography{paper}

\end{document}